\documentclass[prl,10pt,twocolumn]{revtex4}

\usepackage{graphicx}
\usepackage{amsmath,amssymb}
\usepackage{colordvi,rotating}

\newcommand{\be}{\begin{equation}}
\newcommand{\ee}{\end{equation}}
\newcommand{\ba}{\begin{eqnarray}}
\newcommand{\ea}{\end{eqnarray}}
\newcommand{\n}{\nonumber}

\begin{document}
\title{Time-frequency Domain Analogues of Phase
Space Sub-Planck Structures}
\author{Ludmi{\l}a Praxmeyer}
\affiliation{Theoretical Physics Division,
 Sofia University, James Bourchier 5 blvd,
1164 Sofia, Bulgaria}

\author{Piotr Wasylczyk}
\author{Czes{\l}aw Radzewicz}\affiliation{Institute of
 Experimental
Physics, Warsaw University, ul. Ho\.za 69, 00--681 Warsaw, Poland}

 \author{Krzysztof W\'odkiewicz}\affiliation{Institute of Theoretical
Physics, Warsaw University, ul. Ho\.za 69, 00--681 Warsaw,
Poland\\Department of Physics and Astronomy, University of New
Mexico, Albuquerque NM 87131, USA}

\begin{abstract}
We present experimental data of the frequency resolved optical
gating (FROG) measurements of light pulses revealing interference
features which correspond to sub-Planck structures in phase space. For
superpositions of pulses a small, sub-Fourier shift in the carrier
frequency leads to a state orthogonal to the initial one, although
in the representation of standard time--frequency distributions
these states seem to have a non-vanishing overlap.
\end{abstract}

\maketitle

It was shown by Zurek \cite{zurek2001} that sub-Planck structures in
phase space - an unexpected sign of quantum interference - play a
surprisingly important role in {the distinguishability of quantum
states. A sub-Planck phase space shift applied to superposition of
coherent states is sufficient to produce a state which is orthogonal
to the unshifted one.} {This seems counter-intuitive because such
superpositions of coherent states overlap significantly.
Nevertheless, a very small shift causes that the original and shifted
superpositions are orthogonal
to each other, which makes them - at least in principle -
distinguishable.} The effect was originally studied for a
superposition of four coherent states \cite{zurek2001} and then in
\cite{thesis,chirp} it was shown that superpositions of just two
coherent states lead to a similar result.

 { Both
in classical optics and quantum mechanics the linear superposition
principle is the basis of all interference phenomena. Thus, it
should not surprise one that if quantum wave packets are  replaced
 by light pulses, the effects similar to sub-Planck
structures i.e. sub-Fourier structures should be observed. In this
letter an experimental realization of a time-frequency version of
this phenomenon is reported.}

Instead of superpositions of coherent states in phase space, optical
fields in the form of coherent{ superpositions} of pulses are used
and the FROG spectrogram is recorded. A specific cross-{section} of
this spectrogram represents a scalar product of the measured field
and the field with the same envelope but a shifted carrier
frequency. We show that for fields that have the form of {a}
superposition of two pulses this scalar product is an oscillating
function of the frequency shift. Moreover, zeros of this scalar
product are spaced by {sub-Fourier} distances in the scale of the
superposed pulses i.e.{,} the distance between them is smaller than
the single pulse spectral width.


During the last decade FROG (Frequency Resolved Optical Gating) has
become a standard method for reconstruction of the amplitude and
phase of ultrashort light pulses \cite{frog1993,frog1994,frog_book}.
In the second harmonic version of this technique (SH FROG) a pulse
to be measured is split and its two mutually delayed replicas are
overlapped in a nonlinear optical crystal. The sum frequency signal
generated in the crystal is spectrally resolved and recorded for
different time delays $\tau$ . The resulting time--frequency map has
the form \ba\label{frogSHG} { I_{FROG}(\tau,\omega)}=\biggl|\int
dt\, { E(t-\tau)E(t)e^{i\omega t}}\biggl|^2\,, \ea and its
cross-section for zero delay reads
\begin{eqnarray}\label{overlap}
 I_{FROG}(0,\omega)=\biggl|\int\!\! dt\,
 E(t)E(t)e^{i\omega t}\biggl|^2.
\end{eqnarray}

Now, consider a pulsed electric field with a real envelope
${\cal{A}}(t)$ and a linear phase
 \ba E(t)={\cal{A}}(t)e^{-i\omega_c
t},\label{field} \ea where $\omega_c$ denotes  a carrier frequency
of the pulse. The absolute value squared of the scalar product of
this field
 and a field with the same envelope but a carrier
frequency shifted by
$\delta$, i.e.
$E(t)e^{i\delta t}$, is given by
\begin{equation}\label{over}
|\langle E(t)|E(t)e^{i\delta t}\rangle|^2= \left|\int dt
E^{\ast}(t)E(t)e^{i\delta t}\right|^2.
\end{equation}
For transform-limited pulses of a given carrier frequency $\omega_c$
the form of
 $I_{FROG}(0,\delta+2\omega_c)$ is the same as the one given by
Eq.(\ref{over}). In other words, the $\tau=0$ cross-section of the
FROG map measured at $2\omega_c+\delta$ is equal to an overlap of
the field $E(t)$ and the same field shifted in frequency by
$\delta$, i.e. $E(t)e^{i\delta t}$.

For example, when two Gaussian pulses characterized by their carrier
frequency $\omega_c$, dispersion $\sigma$ and time separation
$2t_0$ are superposed:
 \ba
E^{{sup}}(t)=\left(e^{-\frac{(t-t_0)^2}{4\sigma^2}}+
e^{-\frac{(t+t_0)^2}{4\sigma^2}}\right)e^{-i\omega_c t}, \label{sup}
\ea the scalar product (\ref{over}) becomes \ba |\langle
E^{sup}(t)|E^{sup}(t)e^{i\delta t}\rangle |^2=8\pi^2
\sigma^2 e^{-\delta^2\sigma^2} \times\n\\
\left[\cos(\delta t_0) +
e^{-\frac{t_0^2}{2\sigma^2}}\right]^2\!\!\!,\label{cross0} \ea
while the $\tau=0$ cross-section of the corresponding {FROG} spectrogram is
\ba I_{FROG}^{sup}(0,\delta)=8\pi^2 \sigma^2
e^{-(\delta-2\omega_c)^2\sigma^2}\times\n\\
\left[\cos((\delta -2\omega_c) t_0) +
e^{-\frac{t_0^2}{2\sigma^2}}\right]^2.\label{cross}
\ea
Formula (\ref{cross0}) can be derived from (\ref{cross}) after
substituting
$\delta\rightarrow \delta+2\omega_c$.
 For $t_0\neq 0$ Eqs{.} (\ref{cross0}) {and} (\ref{cross})
 have an infinite number of
nearly equally spaced zeros. {This  means that a superposition of
two Gaussian pulses is orthogonal to similar superpositions with
appropriately shifted carrier frequencies}. Moreover, the smallest
of these leading--to--zero--overlap shifts is given approximately by
$\frac{\pi}{2 t_0}$, which for sufficiently large separations $2t_0$
becomes sub-Fourier in a single pulse scale. {These sub-Fourier
shifts correspond directly to the sub-Planck shifts leading to
orthogonality of superpositions of two (or four) coherent quantum
wave packets} \cite{zurek2001,thesis}.

\begin{figure}[h]
\begin{center}
$\!\!\!$\includegraphics[scale=3]{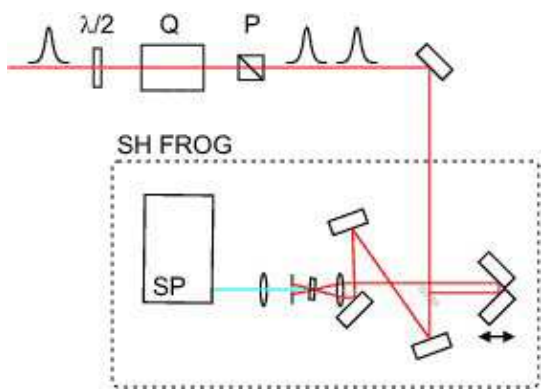}
\end{center}
\caption{{\footnotesize{ The experimental setup. Half waveplate $\lambda$/2, 10 mm crystaline
quartz plate Q and the polarizer P are used to generate the double
pulse. SH FROG is a standard second harmonic FROG with 100 $\mu$m
type I BBO crystal and linear CCD array spectrometer.} }}\label{setup}
\end{figure}

The experimental setup is presented in Fig.\ref{setup}. The pulse
source was a home-built Ti:Sapphire oscillator delivering 50 fs
(FWHM) pulses centered at 780 nm with a 80 MHz repetition rate. To
generate a pulse pair with a highly stable delay we used a half
waveplate, a crystalline quartz block 10 mm thick (optic axis
parallel to the input surface) and a polarizer.
The pulse polarization was first rotated causing the pulse to
split into two replicas of mutually perpendicular polarizations
and similar energies in the quartz crystal. When the polarizer was
set at 45 degrees with respect to the optic axis a pair of delayed
pulses with parallel polarizations emerged. The group delay
between 780 nm pulses propagating as ordinary and extraordinary
rays is approximately
 30 fs/mm in crystalline quartz.  The total delay was fine--tuned by
tilting the quartz block.  The standard FROG apparatus consisted of
a dielectric beamsplitter, a delay line with a stepper motor-driven
translation stage (8 fs/step), a 100 $\mu$m type I BBO crystal for
second harmonic generation and a specrometer with a linear CCD array
of 2048 pixels providing resolution better than 0.5 nm (Ocean
Optics, USB 2000).

\begin{figure}[h]
\begin{center}
$\!\!\!${\small a)}\includegraphics[scale=4]{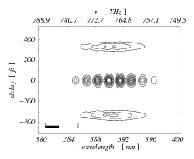}
\\$\;$\\$\!\!\!${\small
b)}\includegraphics[scale=4]{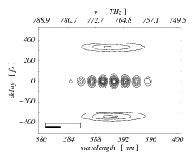} 
\end{center}
\caption{{\footnotesize{FROG maps measured for superpositions of two pulses
with slightly
different separation distances $2t_0$ between the pulses: a)
$2t_0=305\,fs\;$; b)
$2t_0=309\,fs$. Black rectangles represent minimal uncertainty relation areas
of a single pulse, white ones give the corresponding FWHM widths.}
}}\label{fig2}
\end{figure}
\begin{figure}[h]
\begin{center}
$\quad$\\$\quad$\\
{\small a)}
\includegraphics[scale=4]{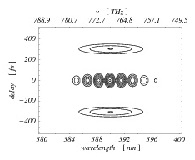}
\\ $\quad$\\$\quad$\\
{\small b)}\includegraphics[scale=4]{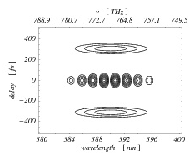}
\end{center}
\caption{{\footnotesize{Numerically calculated
FROG maps corresponding to experiment results presented in Fig.
\ref{fig2}. } }}\label{fig3}
\end{figure}

\begin{figure}[h]
\begin{center}
\begin{picture}(0,0)(35,10)
\put(100,0){\makebox(0,0){\scriptsize{$2t_0=305$ fs }}}
\put(100,-78){\makebox(0,0){\scriptsize{$2t_0=309$ fs}}}
\end{picture}
{\small
a)}\includegraphics[scale=3.]{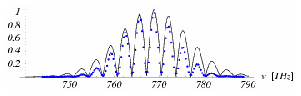}
\\
{\small
b)}\includegraphics[scale=3.]{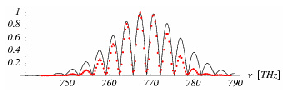}
\caption{{\footnotesize{The $\tau=0$ cross-sections of FROG maps from Fig.
\ref{fig2} a), b). Experimental data are denoted by dots and lines present
the corresponding numerically calculated curves. The relative
phase between the pulses
 was the only free parameter of the fits.}
}}\label{fig4}
\end{center}
\end{figure}

Figures {\ref{fig2}a and \ref{fig2}b} present {the} FROG
spectrograms measured for {a} superposition of two pulses with the
separation $2t_0$ of $305$ fs and $309$ fs, respectively.
Numerically calculated maps for pulse pairs with parameters
corresponding to the measurements are presented in Figs.
{\ref{fig3}a and \ref{fig3}b}. The black rectangles at the
contour-plots show minimal uncertainty relation areas of a single
pulse, $4\pi\triangle t \triangle \nu =1$, the white ones give the corresponding
FWHM widths. Figures {\ref{fig4}a and \ref{fig4}b} show $\tau=0$
cross-sections of the measured spectrograms together with their
theoretical fits.

The measured time dispersion $\triangle t=\sigma$  of the superposed pulses was $20.1\pm 0.5$
fs (i.e $47.3\pm 1.0$ fs FWHM) which imposes a $4.0\pm 0.1$ THz Fourier limit on the
frequency dispersion $\triangle \nu$ (or, equivalently, $9.3\pm 0.2$
THz FWHM). The zeros of the oscillating structures seen at Fig. \ref{fig4} are less then $3.3$
THz apart, which is below the Fourier limit imposed by any of the
single pulses. These zeros correspond to carrier frequencies shifts
for which pairs of pulses have a vanishing overlap. Such a shift can be called a sub-Fourier, although, the Fourier/Heisenberg uncertainty relation is certainly not
violated for the state under study. The smaller is the change of the
carrier frequency leading to the zero overlap, the larger the
separation distance $2t_0$ between the pulses.

It is instructive to study {the} effect demonstrated above
using standard time-frequency distribution functions. The Wigner
distribution  \cite{wigner1932, cohen1995}, which is especially useful for description of classical and
quantum interference \cite{kwgh1998, Dragoman2001},  for a pulse with an
electric field $E(t)$ is defined as
\begin{equation}
\label{wigfunc} W_{E} (t,\omega) = \int \frac{d s}{2\pi} \;
E^{\ast}\left( t+\frac{s}{2} \right) \; e^{i\omega s} \; E\left(
t-\frac{s}{2} \right).
\end{equation}
An important property of the Wigner function that links the Wigner
function of two fields $E_{1}(t)$, $E_{2}(t)$ and their scalar
product,
 is given by the Moyal formula:
\ba
|\langle E_1|E_2\rangle|^2= 2\pi \!\int\! dt\! \int\! d\omega\,
W_{E_1}(t,\omega) W_{E_2}{(}t,\omega).\label{Moyal}
\ea

\begin{center}
\begin{figure}[h]
{\small a)}\includegraphics[scale=1.8]{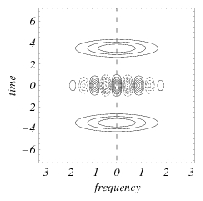}
$\quad${\small
b)}\includegraphics[scale=1.8]{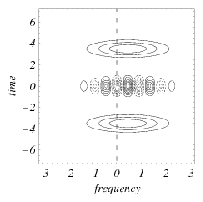}
\caption{\footnotesize  The  Wigner function calculated for a)
two Gaussian pulses, $E(t)= e^{-(t-t_0)^2}+
 e^{-(t+t_0)^2}$, $t_0=3.5$;  b) a similar pair with the carrier frequency
shifted by
$\frac{\pi}{2t_0}$. Plots a) and b) correspond to  mutually
orthogonal pairs of pulses. Dotted contours denote negative values
of the Wigner function.}\label{fig12}
\end{figure}
\end{center}

Figure {\ref{fig12}a} presents a contour plot of the Wigner function
for the superposition of two Gaussian pulses, Eq. (\ref{sup}), with
the separation $2t_0=7$ and $\sigma=1/2$. The second contour plot
(Fig. {\ref{fig12}b}) corresponds to the Wigner function of the same
pair of pulses but with the carrier frequencies shifted by
$\pi/2t_0$. The calculated scalar product of {these} two superpositions is zero.
In the Wigner representation, the vanishing of the scalar product
may be interpreted as an interference effect: the interference
fringes in Fig.{\ref{fig12}b} are shifted by half of the modulation
period with respect to those in Fig.{\ref{fig12}a}. This shift
causes that in the scalar product calculated according to Eq.
(\ref{Moyal}), a negative contribution from the interference terms
{cancels exactly the positive contribution} from {the} Gaussian
peaks.

\begin{center}
\begin{figure}[ht]
{\small a)}\includegraphics[scale=1.8]{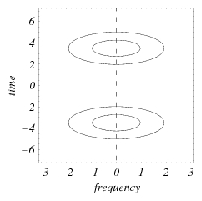}
$\quad${\small
b)}\includegraphics[scale=1.8]{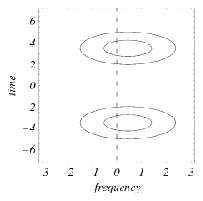}
\caption{\footnotesize The Husimi function calculated for
a) two  Gaussian pulses, $E(t)= e^{-(t-t_0)^2}+
 e^{-(t+t_0)^2}$, $t_0=3.5$;
b) similar pair with the carrier frequency shifted by
$\frac{\pi}{2t_0}$.
Even though the two plots appear very similar the scalar product
of these two pairs vanishes.}\label{fig11}
\end{figure}
\end{center}

Another widely used time-frequency distribution is the Husimi
function {(or the Glauber $Q$-function) }\cite{Husimi}. Figures
{\ref{fig11}a} and  {\ref{fig11}b} present the Husimi function
calculated for the state described by Eq.(\ref{sup}) with {the same
parameters as those used in the evaluation of the Wigner function}.
The plots presented in Figs {\ref{fig11}a}, {\ref{fig11}b} are
 identical except for a small frequency shift. The fact that the scalar product of the
corresponding fields vanishes is rather surprising -- at least until
one recalls that the Moyal formula {given by} Eq.(\ref{Moyal}), is
not applicable to the Husimi function (see Fig. \ref{fig12}).

%

To conclude: we have demonstrated the presence of sub-Fourier
structures in the measured time--frequency representation of light
pulses. These structures are classical counterparts of sub-Planck
features in the phase space.  We have shown that two
pairs of pulses displaced by a small, sub-Fourier shift of the
carrier frequency may be mutually orthogonal even if they seem to
have a non-vanishing overlap in the time-frequency representation of
some commonly used quasi--distributions.

The experiments were performed in the KL FAMO laboratory in Torun,
Poland. This research was partially supported by Polish MNiSW
Grant No. 1 P03B 137 30 and European Union's Transfer of Knowledge
project CAMEL (Grant No. MTKD-CT-2004-014427)
{

\end{document}